# The Role of Solvents in the Formation of Methylammonium Lead Triiodide Perovskite

Junke Jiang,[1,2] José Manuel Vicent-Luna,[1,2] and Shuxia Tao,*[1,2]

[1]Materials Simulation and Modelling, Department of Applied Physics, Eindhoven University of Technology, 5600MB Eindhoven, The Netherlands

[2]Center for Computational Energy Research, Department of Applied Physics, Eindhoven University of Technology, Eindhoven 5600 MB, The Netherlands

**ABSTRACT**

Metal halide perovskites (MHPs) are gaining increasing attention as low-cost, high-performance semiconductors for optoelectronics. In particular, their solution processing is compatible with the large-scale manufacturing of thin-film devices, including solar cells and light-emitting diodes. Understanding the coordination chemistry in precursor-solvent solution and atomistic mechanisms of film formation is of great importance for optimizing the optoelectronic properties of the final films. Using the methylammonium lead triiodide ($MAPbI_3$) as an example, we study the complex evolution of the molecular species from the solution to the initial stage of the crystallization by using a combination of density functional theory (DFT) calculations and *ab-initio* molecular dynamics (AIMD) simulations. We focus on the widely employed solvents DMSO and DMF, analyze the structures and energies of the iodoplumbate complexes in the form of simple complex of $[PbI_mL_n]^{2-m}$ and polymeric iodoplumbates of $([PbI_mL_n]^{2-m})_x$. Based on the calculated formation enthalpies, we propose reaction schemes of $MAPbI_3$ formation in DMSO, DMF and DMSO-DMF binary solvent and explain the advantages of the binary solvent. We highlight the important role of NH¨O hydrogen bonds in the formation of iodoplumbates monomers. Our calculations indicate unbalanced reaction energies at several elementary reaction steps in either DMF (formation of $[PbI_4L_n]^{2-}$ being highly favourable) or DMSO (formation of $[PbI_5L_n]^{3-}$ being retarded). Mixing a small amount of DMSO in DMF gives rise to a better balance in the energies and, therefore, potentially better equilibria in the overall crystallization process and better quality of the final perovskite films.

**INTRODUCTION**

Metal halide perovskites (MHPs) are among the most studied semiconducting materials in the last decade due to their promising optoelectronic properties, making highly efficient solar cells and light-emitting diodes.[1-4] The impressive progress in achieved power conversion efficiency (PCE) has consolidated their popularity among photovoltaic materials.[5-8] The MHPs are made by simple solution-processed deposition techniques by mixing a metal-halide precursor with a halide salt in a solution. For instance, methylammonium iodide (MAI) is added to a lead halide ($PbI_2$) solution in coordinating solvents, such as dimethylformamide (DMF),[9, 10] dimethyl sulfoxide (DMSO),[9, 11] and N-methyl-2-pyrrolidone (NMP).[12, 13] From the mixture, colloidal particles first form and then produce a MHPs film by solvent evaporation. However, in contrast to the simplicity of the synthesis, the physicochemical processes during the synthesis are very complex. The synthesis involves the solvation/de-solvation and complexation equilibria of all participating species at solvents-precursors, nucleation, and it evolves various intermediate phases in the two-step method,[11, 14, 15] and/or iodoplumbate complexes in the one-step method.[10, 16, 17]

Using $MAPbI_3$ as an example, various complexes could be formed in a solution, depending on the type and the concentration of ions ($MA^+$ cations and $I^-$ anions) or solvents (L) near the Pb metal ions. When $I^-$ anions together with $MA^+$ cations are added to the lead iodide ($PbI_2$) solution, competition in binding occurs between $I^-$ and solvent molecules to the $Pb^{2+}$ cation. The $I^-$ anion and solvent molecules coordinate around the central $Pb^{2+}$ cation, forming iodoplumbate complexes, i.e., $[PbI_2L_n]$, $[PbI_3L_n]^{1-}$, $[PbI_4L_n]^{2-}$, $[PbI_5L_n]^{3-}$, and $[PbI_6L_n]^{4-}$.[14, 18-22] During the iodoplumbates formation, a small amount of $I^-$ is adequate to replace the solvent molecules, thus induce the formation of iodoplumbate species.[18, 20] Indeed, iodoplumbate complexes, such as $[PbI_3L_n]^-$ and $[PbI_4L_n]^{2-}$ can be stabilized in the presence of large excess iodide ions.[18, 19] The high iodide coordinated iodoplumbate, such as $[PbI_5L_n]^{3-}$ and $[PbI_6L_n]^{4-}$ are less often directly observed in experiments. However, their presence cannot be excluded, and it is speculated to play an important role in the final steps of the formation of MHPs.[14, 20-22]

Besides the simple iodoplumbates discussed above, polymeric iodoplumbates was proposed to be formed, using the simple complexes, such as $[PbI_2L_n]$, $[PbI_3L_n]^{1-}$, or $[PbI_4L_n]^{2-}$ monomers as building blocks.[18, 23] The atomistic structure, electronic, and optical property of the polymeric species during the MHPs formation has been extensively investigated.[17, 18, 23] Oleksandra and co-workers [18, 23] proposed the formation of polynuclear, i.e. formation of $([PbI_mL_n]^{2-m})_x$ from $[PbI_2L_4]$, is essential in the formation of MHPs. It has been also proposed that the polymeric iodoplumbates in the form of $([PbI_mL_n]^{2-m})_x$ most likely occur when using a high concentration of lead-iodide in the precursor solutions.[18, 23] However, an atomistic scale evolution of these molecular complexes remains unclear, and the crystallization pathways at various synthesizing conditions remain highly debated.

Another important factor that impacts the coordination chemistry of MHPs are the solvents.[24, 25] The most commonly used solvents are DMF [9, 10] and DMSO.[9, 11] The usage of binary solvent, one as the basic solvent and another as co-solvent, i.e. coordinative additives, is also widely used in the synthesis of MHPs.[13, 14, 26-28] Among various co-solvents, DMSO the most used ones in combination with DMF.[14, 27, 29] It has been proposed from experiments that the use of DMSO-DMF binary solvent could retards the crystallization by forming MAI-$PbI_2$-DMSO complexes, and thus slowing down the nucleation rate of MHPs and therefore leads to higher quality films.[29-31] However, the atomistic origin of such effects and evolution of the relevant complexes involved in the crystallization is challenging to obtain in experiments due to the limitations in the spatio temporal resolutions.

The atomistic simulation is a powerful tool to investigate the fundamental aspects of chemical processes during crystallization at an atomic scale. However, due to the complexity of the species involved, theoretical studies are only emerging. Clancy *et al.*[32] investigated the relative stability of precursors in solution, and showed that the Mayer bond order could predict the solubility of the lead halide precursors in the solvent. Rothlisberger *et al.* [33] used metadynamics to investigate the nucleation of MHPs, mimicking a one-step synthesis route. Recently, Angelis *et al.*[19, 21] have studied the formation of high-I-coordinated iodoplumbates. They concluded that a large amount of MAI is needed for the formation of iodine-rich iodoplumbates.[19] This implies the difficulty in the formation of $[PbI_5L_n]^{3-}$ and $[PbI_6L_n]^{4-}$, which are considered necessary for the formation of high-quality perovskite films.[9, 22, 34-38] To our knowledge, the conversion from the low-I-coordinated iodoplumbates (i.e., $PbI_2$) to high-I-coordinated iodoplumbates and their conversion to polymeric iodoplumbates is not systematically studied yet. Also, the highly dynamic formation process of the iodoplumbate complexes can be significantly affected by the A-cations as well as the type of solvents. The interplay of this process has not been deeply investigated and remains unclear.

In this work, using $MAPbI_3$ and two commonly employed solvents (DMF and DMSO), we investigate the evolution of the iodoplumbates in solutions using a combination of DFT and AIMD simulations. We obtain information on structures and reaction energies of elemental reaction steps from low to high-I-coordinated complexes and investigate how solvents impact the overall reaction pathways. We start with the commonly used solvents, DMF and DMSO and study the impact of the co-solvent DMSO in DMF, by constructing a binary DMF-DMSO model. By analyzing the changes in the structures and energetics of elementary steps, we discuss the advantages of the binary solvent. Our work provides an atomistic understanding of the first steps of the crystallization of MHPs and forms a basis for further investigation of more complex perovskite systems or longer time scale simulations to model complete crystallization processes.

## METHODS

### Density functional theory calculations

All DFT calculations were performed in the Kohn–Sham framework, using the Amsterdam Density Functional package (version ADF 2019.302).[39, 40] Geometry optimizations were done using the Perdew-Burke-Ernzerhof (PBE) functional.[41] A combination of Herman-Skillman numerical atomic orbitals (NAOs) with Slater type Triple-zeta (TZP) basis sets were used for heavy atoms (Pb and I), whereas for C, N, H, O, and S atoms, a Double-zeta (DZP) basis sets were used. The cores 1s−4d, 1s−4p, and 1s−2s were kept frozen, respectively, for Pb, I, and S, to speed up the calculations. Vibrational frequencies were obtained with the optimized configurations through numerical differentiation of the analytical gradient. The harmonic level approximation was applied to ensure that the optimized structures are located at global or local minima on the potential energy surface.[42-44] Figure S1 shows a representative example of the distance dependence on the interaction energies between $MAPbI_3$ complexes and DMF solvents. The enthalpies were calculated at 298.15 K and 1 atm from the bond energies and vibrational frequencies by using a standard thermochemistry relation as described by Bickelhaupt *et al.*[45, 46]. The solvents were simulated with both implicit solvation models (the COSMO dielectric continuum solvation scheme) [47] and explicit solvent molecules. The hybrid implicit-explicit models were chosen as it was proven to be useful for better describing the precursor chemistry of halide perovskites in previous publications.[19, 21, 32, 48]

### *Ab Initio* Molecular Dynamics Simulation

AIMD simulations were also performed to study the dynamics of precursor-solvent interactions by using the CP2K code.[49, 50] The QUICKSTEP, the electronic structure part of CP2K, uses the combined Gaussian and plane-wave (GPW) method to calculate forces and energies. The GPW method is based on the Kohn-Sham formulation of DFT and employs a hybrid scheme combining Gaussian and plane wave functions. The Born-Oppenheimer molecular dynamics (BOMD) simulations were executed by using a canonical ensemble (NVT) with a Nosé–Hoover thermostat [51] at temperature 353 K (typical temperature for the spin-coating procedure in experiments).[13, 28, 52-54] All simulations made use of DFT at the PBE level with double-zeta basis sets (DZVP-MOLOPT for Pb, I, S, O, C, N, H) [55] and Goedecker-Teter-Hutter (GTH) pseudopotentials [56] with a 500 Ry density cut-off. A single *k*-point sampled at the Γ point was used to speed up the computation. The total simulation time of each calculation is 30 ps with a time step of 1 fs.

### Structural models

To calculate the formation enthalpy of the iodoplumbates with increased iodide coordination, models $[PbI_mL_n]^{2-m}$ (m = 2, 3, 4, 5, 6; n = 6) for simple complex and $[(PbI_mL_n)^{2-m}]_2$ (m = 2, 3, 4; n = 6-m) for polymeric iodoplumbates were used: m is the number of coordinated $I^-$, L is the solvent molecules, and n is the number of solvents. The scheme of the formation of both simple and polymeric iodoplumbates is shown in Figure 1. In order to ensure charge neutrality, the number of $MA^+$ cations included in the models is equal to the number of the access $I^-$ ions (m-2 for simple complex and 2m-4 for polymeric iodoplumbates). For AIMD simulations, the homogeneous mixture of ions with fixed loading: 6 $[PbI_3]^-$, 6 $MA^+$, and 18 solvents were constructed with lattice parameters of 14.5 Å. The calculated molar concentration of MAI / $PbI_2$ blend in DMF is equivalent to 1.82 M, according to the second-order polynomial fitting of the results obtained by Zhang *et al.* [57], which is similar to the experiment concentration of 1.64 M (see Figure S2). In the DMF-DMSO binary solvents model for DFT calculations, we incorporated 1 DMSO in DMF-DMSO binary solvent models (DMSO:DMF = 1:5) and DMF is modeled by using the hybrid implicit-explicit solvent model. While for AIMD simulations of binary solvents model, we included 2 DMSO out of how many 18, resulting in the ratio of DMSO:DMF = 1:8. The atomic structures and key parameters (donor number and boiling point) of DMSO and DMF are shown in Table S1.

## RESULTS AND DISCUSSION

### Formation enthalpies of iodoplumbates

The formation of iodoplumbate complexes is an important step for the MHPs crystallization. To gain insights into the formation of these iodoplumbates, we considered from low to high I-coordination, i.e. from $[PbI_2L_6]$ to $[PbI_6L_6]^{4-}$ in this study. As illustrated in Figure 1 (top), the conversion from low I-coordinated iodoplumbate (i.e., $[PbI_2L_6]$) to high I-coordinated iodoplumbate ($[PbI_3L_6]^{1-}$ $[PbI_4L_6]^{2-}$, $[PbI_5L_6]^{3-}$, and $[PbI_6L_6]^{4-}$) involves reactions that require or release energy. It should be noted that the negative charge on the coordination sphere due to the access $I^-$ is compensated by adding positively charged $MA^+$ ions in our simulations. The formation enthalpy of each step is calculated using the Equations 1 - 4 as following:

$$[PbI_2L_6] + MAI \rightarrow [PbI_3L_6]^{1-} + MA^+ \quad (1)$$

$$[PbI_3L_6]^{1-} + MA^+ + MAI \rightarrow [PbI_4L_6]^{2-} + 2MA^+ \quad (2)$$

$$[PbI_4L_6]^{2-} + 2MA^+ + MAI \rightarrow [PbI_5L_6]^{3-} + 3MA^+ \quad (3)$$

$$[PbI_5L_6]^{3-} + 3MA^+ + MAI \rightarrow [PbI_6L_6]^{4-} + 4MA^+ \quad (4)$$

To study the formation of polymeric iodoplumbate, we focus on the investigation of the formation of iodoplumbates dimers $[PbI_2L_4]_2$, $([PbI_3L_3]^{1-})_2$, and $([PbI_4L_2]^{2-})_2$ from the simple iodoplumbates, $[PbI_2L_6]$, $[PbI_3L_3]^{1-}$, and $[PbI_4L_2]^{2-}$, respectively. These three iodoplumbates are commonly detected complexes in experiments.[18, 20, 22, 23] The formation enthalpy of the polymeric iodoplumbates is evaluated by using Equations 5 - 7, as following:

$$[PbI_2L_4] + [PbI_2L_4] \rightarrow [PbI_2L_4]_2 \quad (5)$$

$$[PbI_3L_3]^{1-} + [PbI_3L_3]^{1-} \rightarrow ([PbI_3L_3]^{1-})_2 \quad (6)$$

$$[PbI_4L_2]^{2-} + [PbI_4L_2]^{2-} \rightarrow ([PbI_4L_2]^{2-})_2 \quad (7)$$

The formation enthalpies for all steps with respect to different solvents are summarized in Figure 1. A negative formation enthalpy indicates that the corresponding iodoplumbates formation reaction is favourable.

We first look at the step-by-step formation of iodoplumbate complexes from low to high I-coordination. For DMF, the reaction of all steps is favourable, evidenced by the negative formation enthalpies. In particular, the formation enthalpy from $[PbI_3L_6]^{1-}$ to $[PbI_4L_6]^{2-}$ is -0.72 eV, which is much more negative (about 0.5 eV) than other steps, indicating a rapid formation of $[PbI_4L_6]^{2-}$. For DMSO, the formation enthalpies are all negative (in the range of about -0.4 to -0.5 eV), except for the step from $[PbI_4L_6]^{2-}$ to $[PbI_5L_6]^{3-}$, which is close to 0 (-0.07 eV). This indicates that the formation of $[PbI_5L_6]^{3-}$ becomes unfavourable. When mixing a small amount of DMSO in DMF, i.e. using a DMF-DMSO binary solvent, we observe interesting changes in the reaction energetics compared to pure DMF or DMSO, respectively. While the formation enthalpy of all steps are negative, that from $[PbI_3L_6]^{1-}$ to $[PbI_4L_6]^{2-}$ become relatively small, being -0.19 eV, compared to -0.72 eV and -0.48 eV for DMF and DMSO, respectively. This indicates the conversion from $[PbI_3L_6]^{1-}$ to $[PbI_4L_6]^{2-}$ become relatively difficult, i.e., relatively slower coordination reactions. It is worth noting that the conversion of $[PbI_4L_6]^{2-}$ to $[PbI_5L_6]^{3-}$ becomes more likely, evidenced by the much more negative reaction energy (-0.47 eV *vs* -0.21/-0.07 eV) for DMF-DMSO binary compared to DMF or DMSO.

To gain insights into the underlying mechanisms of the significant changes in the reaction energies from $[PbI_3L_6]^{1-}$ to $[PbI_4L_6]^{2-}$ and from $[PbI_4L_6]^{2-}$ to $[PbI_5L_6]^{3-}$ by using DMF-DMSO binary solvent, we analyzed the atomistic structures of the relevant complexes. Neither significant changes nor general trends in the average Pb-I bond or Pb-O bond were found from $[PbI_2L_6]^{1-}$ up to $[PbI_5L_6]^{3-}$ (see Figure S3). An interesting interaction pattern in hydrogen bonds, NH┈O, between N-H bonds of $MA^+$ cation and the O atom of the solvents, can be seen in Figure 2. The number of NH┈O hydrogen bonds in DMF or DMSO increases from $[PbI_2L_6]^{1-}$ to $[PbI_3L_6]^{1-}$ and to $[PbI_4L_6]^{2-}$ but decreases at $[PbI_5L_6]^{3-}$. The increase can be explained by the fact more $MA^+$ are added during the reaction from low to high I-coordinated complexes. Such increase is especially significant with a net increase of 2 NH┈O hydrogen bonds from $[PbI_3L_6]^{1-}$ to $[PbI_4L_6]^{2-}$, explaining the largely negative reaction energies (-0.72/-0.48 eV for DMF/DMSO) compared to the reactions before and after this step. However, the trend is interrupted at $[PbI_5L_6]^{3-}$ and the number of NH┈O decreases. This causes less negative reaction energies (-0.21/-0.07 eV for DMF/DMSO) at this step, because more (weaker) NH┈I hydrogen bonds are formed instead of (stronger) NH┈O bonds in $[PbI_5L_6]^{3-}$ (see Figure S4). In contrast, for DMF-DMSO binary solvents, the number of NH┈O hydrogen bonds does not change from $[PbI_3L_6]^{1-}$ to $[PbI_4L_6]^{2-}$ and then increases from $[PbI_4L_6]^{2-}$ to $[PbI_5L_6]^{3-}$. The consequence of this is the formation energy of $[PbI_4L_6]^{2-}$ become less negative (-0.19 eV *vs* -0.72/-0.48 eV) and that of $[PbI_5L_6]^{3-}$ become more negative (-0.47 eV *vs* -0.21/-0.07 eV). These findings show the importance of A-cation $MA^+$ and the unique role of NH┈O hydrogen bonds in the formation energies of iodoplumbates, a factor that has not been considered in previous studies to the best of our knowledge.

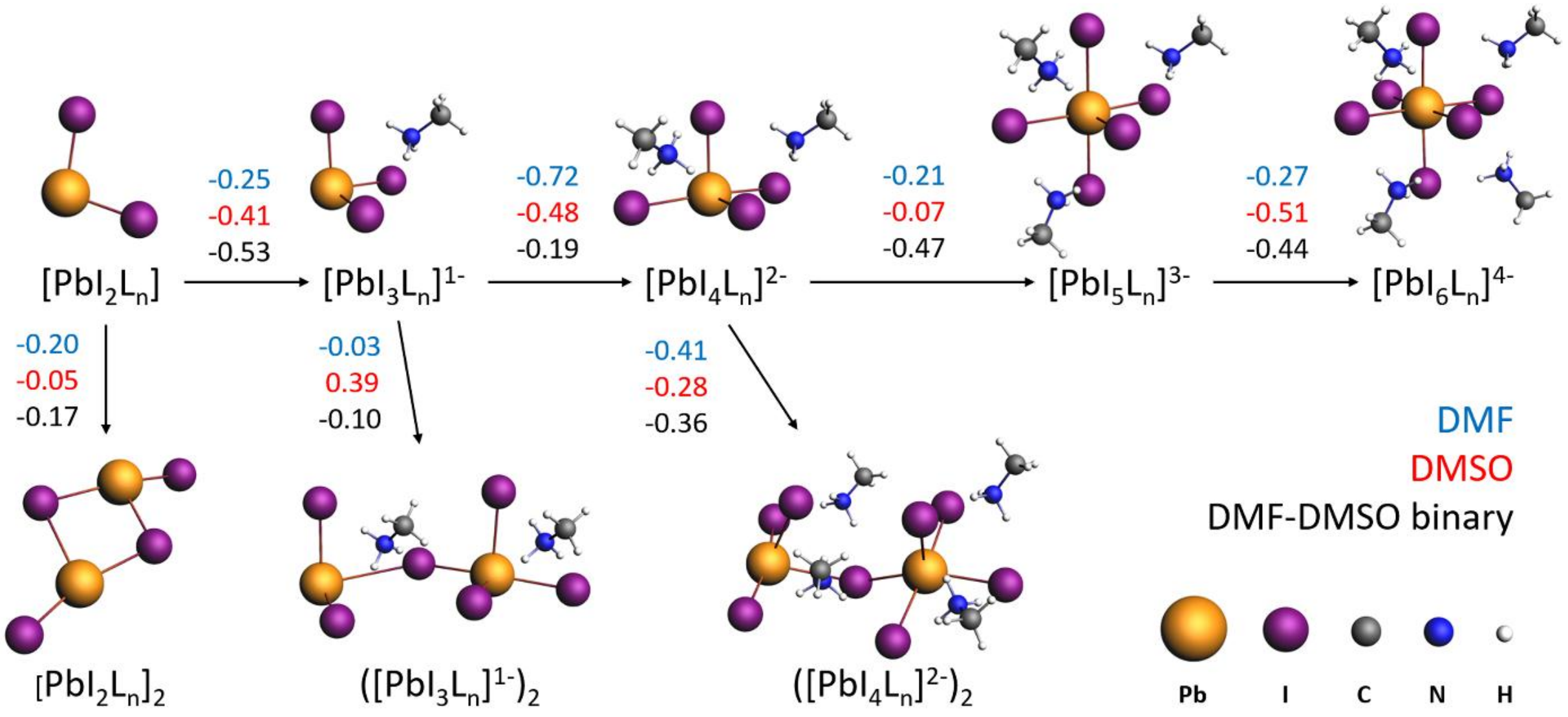

**Figure 1**. Schematic illustration of the iodoplumbate complexes formation: top images are for the simple iodoplumbate complexes, and bottom images are for polymeric iodoplumbate complexes. Formation enthalpies (in eV) of each steps using DMF, DMSO, and DMF-DMSO binary as solvents are shown with values in blue, red and black, respectively. The reaction energies shown in the scheme are calculated by averaging four data points from supporting information in Tables S2, S3, S5, S6, S8, and S9. The calculations were done using a hybrid explicit-implicit solvent model. The structures shown are obtained by using solvent DMF and for clarity, the solvents molecules are not shown in the scheme.

From the energies shown in Figure 1 (bottom), most polymeric complexes can be formed in DMF (evidenced by the negative formation enthalpies) with one exception: the formation energy of $([PbI_3L_3]^{1-})_2$ being almost zero, -0.03 eV. Another important trend is that the energies in DMSO are all more positive than those in DMF, in particular, the formation energy of $([PbI_3L_3]^{1-})_2$ is as high as 0.39 eV. The overall trend can be readily explained by the solvent coordination power of these two solvents with a general trend of DMSO > DMF.[21, 36, 58-62] Indeed, the atomistic structure in Figure S5 show the two $[PbI_3L_3]^{1-}$ monomers cannot or only weakly interact with each other in DMSO because of the strong coordination/interactions of DMSO molecules with $Pb^{2+}$, giving rise to the large positive reaction energy of 0.39 eV. A higher likelihood of forming polynuclear complexes can be expected for solvents with weaker coordination ability, such as DMF. This is because the solvent with weak coordination can be replaced more easily by the $I^-$ in the adjacent monomer (see Figure S5). The DMF-DMSO binary solvent gives values in between those of DMF and DMSO for $[PbI_2L_4]_2$ and $([PbI_4L_2]^{2-})_2$, but brings down the value for $([PbI_3L_3]^{1-})_2$ to below zero (-0.10 eV).

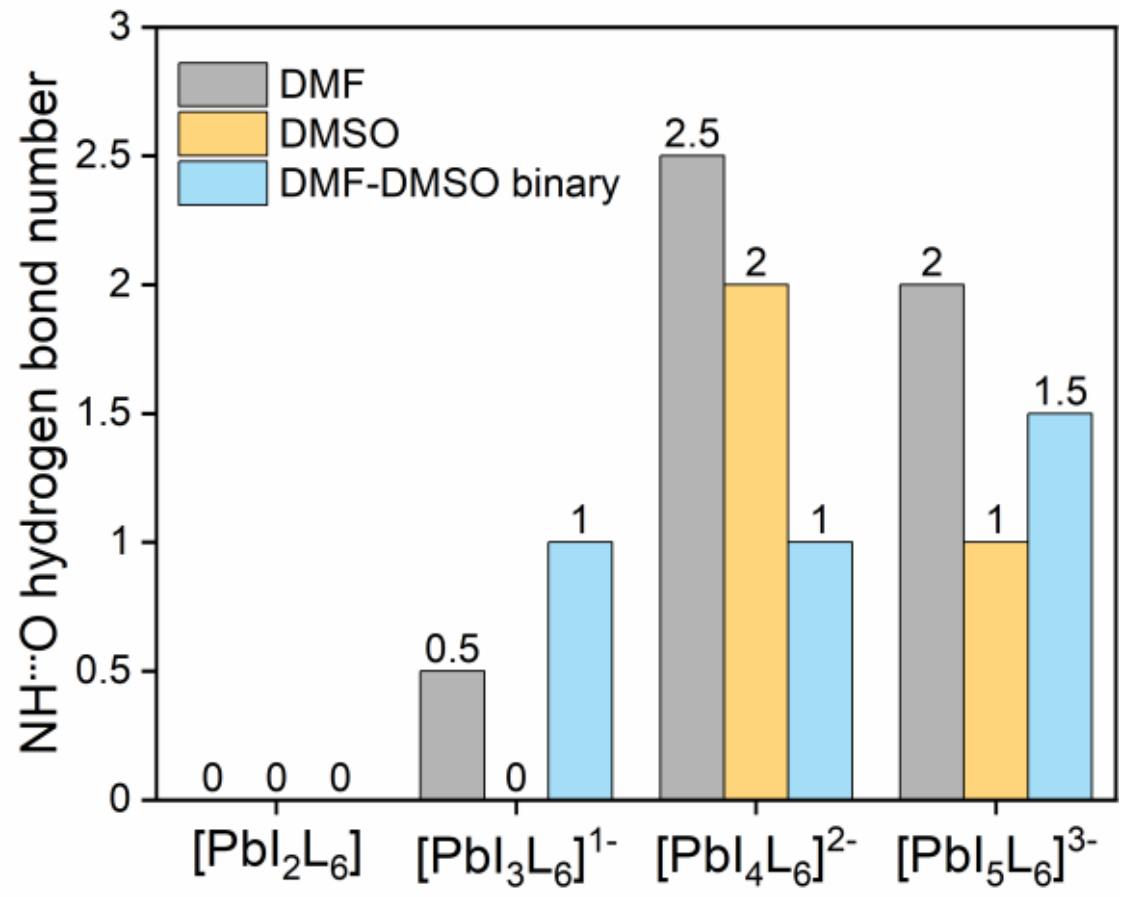

**Figure 2.** Number of NH···O hydrogen bonds for $[PbI_2L_6]$, $[PbI_3L_6]^{1-}$, $[PbI_4L_6]^{2-}$, and $[PbI_5L_6]^{3-}$ in DMSO, DMF, an their mixture. Typically, the length of NH···O hydrogen bonds is shorter than 2.0 Å.[63] The value shown in the figure is averaged from all NH···O hydrogen bonds of two iodoplumbate configurations considered (Table S4, S7, and S10).

From the reactions of the simple and polymeric iodoplumbates, a combined reaction scheme can be constructed in Figure 1 and the effect of the solvents can be analyzed. For DMF, the step-by-step reaction from low to high-I-iodoplumbate are all favourable, especially the formation from $[PbI_3L_6]^{1-}$ to $[PbI_4L_6]^{2-}$, indicating a rapid formation of $[PbI_4L_6]^{2-}$. Polymeric complexes such as $[PbI_2L_n]_2$ and $([PbI_4L_3]^{2-})_2$ can be formed except for the $([PbI_3L_3]^{1-})_2$. The overall energy pathway indicates a rapid conversion from low to high-I-coordinated $PbI_x$ complex. For DMSO, all reaction energies at each step, including the formation of simple and polymeric complex, have become more positive compared to those in DMF. In particular, the step from $[PbI_4L_6]^{2-}$ to $[PbI_5L_6]^{3-}$ become much more difficult with a reaction energy of close to 0 (-0.07 eV). This indicates the formation of high-I-coordinated complex, such as $[PbI_5L_6]^{3-}$ is hindered. When mixing a small amount of DMSO in DMF, the reaction energies of all reaction steps fall nicely in a range of -0.50 to -0.10 eV, slowing down the fast conversion from $[PbI_3L_6]^-$ to $[PbI_4L_6]^{2-}$ in DMF and speeding up the difficult conversion from $[PbI_4L_6]^{2-}$ to $[PbI_4L_6]^{2-}$ in DMSO. Such balanced energetics can potentially lead to better equilibria among all reaction steps and therefore better final perovskite films. Our analysis is in agreement with several experimental findings that point to the fact when mixing DMSO in DMF, the formation of MAI-$PbI_2$-DMSO complexes slows down the overall nucleation rate and therefore leads to higher quality final film.[29, 30]

### Formation dynamics of MHPs with different solvents

We next investigate the dynamical evolution [64-68] of the relevant complexes starting from $[PbI_3L_n]^{1-}$ by using *ab-initio* molecular dynamics (AIMD) simulations. The reason for starting with $[PbI_3L_n]^{1-}$ is twofold: i) for saving computational cost; ii) the formation enthalpies starting from $[PbI_3L_n]^{1-}$ are with a significant difference, especially the value of DMSO are remarkably positive. All simulations were done for 30 ps at a typically spin-coating temperature 353 K.[13, 28, 52-54] The trajectory of the initial 5 ps in each simulation is discarded for the data analysis because of the relatively large temperature and energy oscillation (see Figure S6).

Figure 3 illustrates the snapshots of the beginning and the end of the frames in the AIMD simulations. For the DMF, we observe formation of high-I-coordinated iodoplumbate such as $[PbI_4L_n]^{2-}$ and $[PbI_5L_n]^{3-}$ readily within only 30 ps, and face sharing polymeric complexes, indicating a relatively fast nucleation rate. In contrast, in DMSO, we only observe a small amount of corner-sharing polymeric complexes, with the majority of the species preserving the feature of starting monomer $[PbI_3L_n]^{1-}$. Moving to the DMF-DMSO binary, we find a large block of $PbI_x$ complex that is centered by fully-coordinated iodoplumbates, which is surrounded by some corner-sharing $[PbI_3L_n]^{1-}$. We speculate such complex is a starting point for the formation of 3D perovskite structure (fully coordinated $PbI_x$ complex that are cornered shared with each other). Indeed,

Rothlisberger *et al.*[33, 69] demonstrated a slow transition from edge-sharing octahedra to the corner-sharing octahedra after hundreds of nanoseconds using metadynamics. This explains the reason we do not observe a full transition to 3D perovskites in the time scale of our AIMD simulations. The quantitative trends are however well captured in our MD simulations which corroborate the findings from the thermodynamical analysis presented above.

To zoom in to the atomistic details of the simulations from Figure 3, we plot in Figure 4 the short-range radial distribution function $g(r)$ and cumulative radial distribution functions int[$g(r)$] (coordination number) for Pb-I and Pb-O bonds in different solvents. The $g(r)$ of Pb-I shows that the iodine mainly interacts with Pb at the typical Pb-I bond distance of slightly longer than 3.0 Å, which agrees with the results from Angelis *et al.*[70] by using Car–Parrinello molecular dynamics (CPMD) simulations for the two-step formation of $MAPbI_3$. The int[$g(r)$] of Pb-I shows that the average coordination number of iodine to the Pb is between 3 and 4 for DMSO and DMF, and larger than 4 for the DMF-DMSO binary solvent. This trend is supported by the trend seen in Pb-O bonds, where the number of the Pb-O coordination, being about 0.2, is the smallest in the binary solvent, with DMF being 1 and DMSO being 0.5, respectively. This is in concordance with the NH···O (hydrogen bonds between $MA^+$ and solvents) shown in Figure S7, where the solvents with the highest NH···O coordination leads to the lowest Pb-O coordination and viceversa. This suggests that the mutual interactions between DMF and DMSO and with the $MA^+$ cations affect the formation and molecular structure of the $PbI_x$ complexes. The first peak of the Pb-I and Pb-O g(r) is displaced to slightly longer distances in the case of DMF-DMSO binary solvents (from 3.05 Å to 3.15 Å for Pb-I and from 2.65 Å to 2.75 Å for Pb-O bonds, respectively). This facilitates the $PbI_x$ complexes clusterization as well as the removal of the solvent in further steps of the crystallization process.[29, 30] These findings are in agreement with the formation enthalpies described in Figure 1 and support the use of DMF-DMSO binary solvent for promoting the growth of high I-coordinated iodoplumbates, i.e., the first step of the crystallization process.

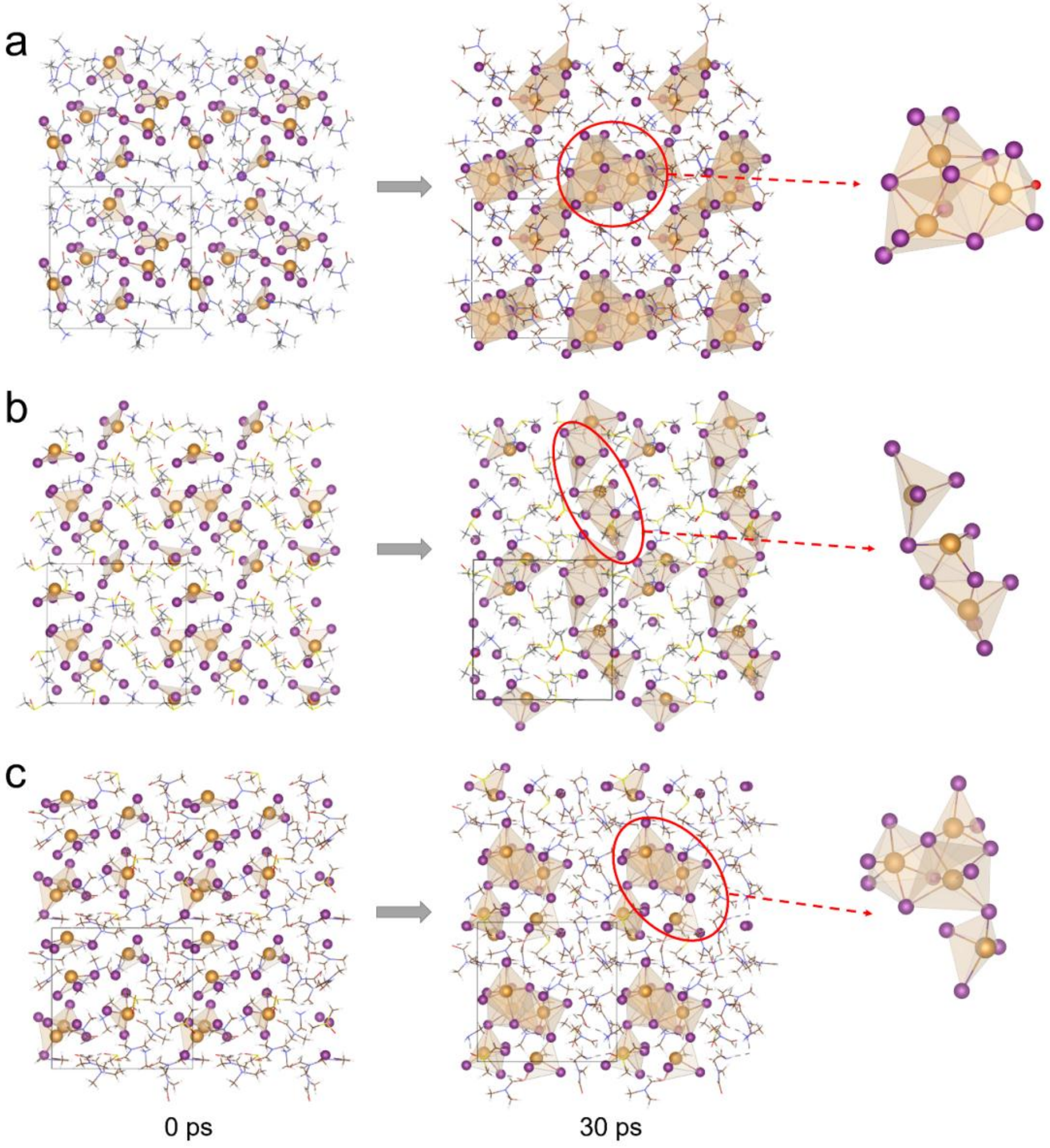

**Figure 3**. Snapshots of the initial and final state of the precursor-solvent mixture during 30 ps AIMD simulation for (a) DMF, (b) DMSO, and (c) DMF-DMSO binary solvents, respectively. The homogeneous mixture of ions contains 6 $[PbI_3]^-$, 6 $MA^+$, and 18 solvents. The snapshots are shown with 2 × 2 supercells, with the unit cell highlighted by the simulation boxes.

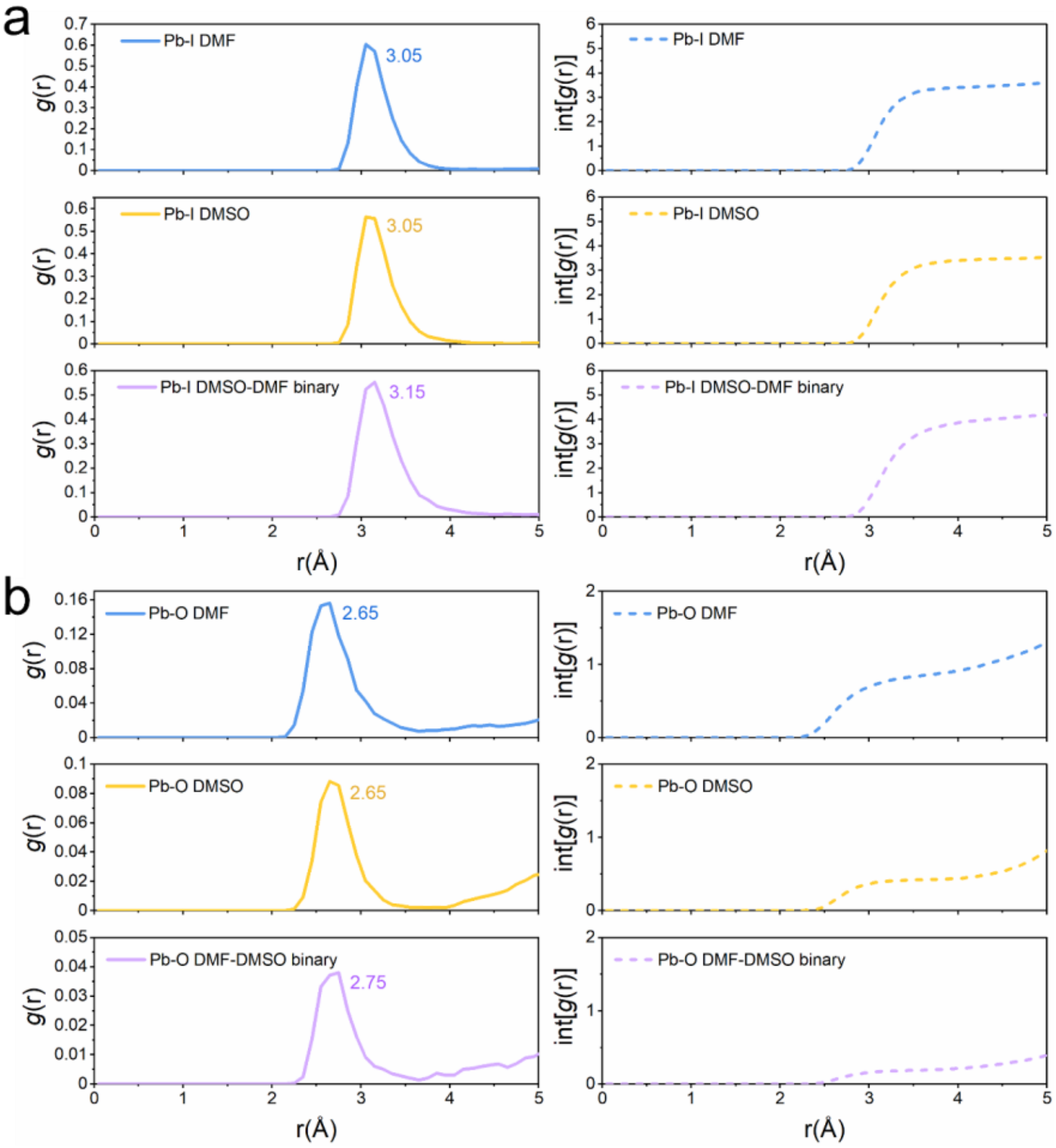

**Figure 4**. Short-range radial distribution function *g*(r) (left) and cumulative radial distribution functions int[*g*(r)] (right) of (a) Pb-I, and (b) Pb-O for DMF, DMSO, and DMF-DMSO binary solvents, respectively.

## CONCLUSIONS

In summary, to understand the coordination chemistry during the formation of MHPs, we study the interactions of perovskite precursors with solvents by combining DFT and AIMD calculations. We construct a reaction scheme for the crystallization of $MAPbI_3$ at the early stage by combining reactions that form simple and polymeric $PbI_x$ complexes and analyze the effect of the commonly used solvents, DMF, DMSO as well as DMF-DMSO binary solvents. For DMF and DMSO, the step-by-step reactions from low to high I-iodoplumbate are all favourable, as well as for $[PbI_2L_n]_2$ and $([PbI_4L_3]^{2-})_2$ polymeric complexes. While DMF favors the formation of low-coordination complexes, from $[PbI_3L_6]^{1-}$ to $[PbI_4L_6]^{2-}$, DMSO hinders reaction of high-coordinated complexes, from $[PbI_4L_6]^{2-}$ to $[PbI_5L_6]^{3-}$. However, when using DMSO as an additive for DMF, the reaction energies are better balanced than for the pure solvents, i.e. the formation of low-coordinated complexes is slower compared to the fast conversion in DMF or DMSO, while the formation of high-coordinated complexes is promoted. Thus, the overall energy pathway is well balanced by mixing a small amount of DMSO in DMF, leading to a better equilibrium in all the reaction steps. This energetic balance in the reactions of iodoplumbates monomers can be explained because of the formation of NH···O hydrogen bonds between the MA cations and the solvents. For the pure DMF and DMSO solvents, the NH···O hydrogen bond number highly increases from $[PbI_3L_6]^{1-}$ to $[PbI_4L_6]^{2-}$, explaining the largely negative reaction energies compared to the reactions before and after this step. However, the trend is interrupted at $[PbI_5L_6]^{3-}$ and the number of NH···O decreases. Interestingly, for DMF-DMSO binary solvents, the number of NH···O hydrogen bonds slightly increases from $[PbI_4L_6]^{2-}$ to $[PbI_5L_6]^{3-}$, causing a more negative reaction energy associated to the formation of high-coordinated $[PbI_5L_6]^{3-}$ complexes.

The dynamical results from AIMD simulations confirmed our predictions of the thermodynamic properties obtained with the DFT calculations. In DMF, while high-I-coordinated $PbI_x$ complex quickly forms, they exhibit face-sharing configuration. For DMSO, we found that the majority of the species remain unreacted. However, the mixture of DMF and DMSO in a binary solvent promotes not only high-I-coordinated iodoplumbates, but also $PbI_x$ complexes with a corner-sharing feature, which can be served as nucleation centers to form 3D perovskites. The structural analysis of the Pb-I and Pb-O bonds is in concordance with the NH¨O hydrogen bonds detailed above. The radial distribution functions and the coordination numbers reveal that the binary DMF-DMSO solvents promote higher Pb-I coordination at the same time that hinder the interaction between the lead atoms and the solvents, which facilitates their removal in further steps of the crystallization process.

Our results provide insights into the role of precursors and solvents in the thermodynamics and kinetics of perovskite formation. The detailed atomistic analysis offers a unique and experimentally inaccessible insight regarding the early stages of the perovskite crystallization and provides an important basis for future work on more complex perovskite compositions and/or a complete crystallization process.

## ASSOCIATED CONTENT

### Supporting Information

The interaction energies of $MAPbI_3$-DMF iodoplumbate complexes as a function of the distance of their center of masses; the correlation between theoretically calculated concentration and practical concentration; atomistic structure, donor number and boiling point of solvents; completed data of all the enthalpy and formation enthalpies for solvents with different configurations; Pb-I and Pb-O bond lengths; the optimized iodoplumbates configurations $[PbI_5L_6]^{3-}$ of DMF and DMSO; the optimized polymeric iodoplumbates (dimer) configurations; temperature flection during AIMD simulation; $g$(r) and int[$g$(r)] of NH¨O for different solvents.

## AUTHOR INFORMATION

### Corresponding Authors

*E-mail: S.X.Tao@tue.nl (Shuxia Tao)

### Notes

There are no conflicts to declare.

## ACKNOWLEDGMENTS

J.J. acknowledges funding by the Computational Sciences for Energy Research (CSER) tenure track program of Shell and NWO (Project No. 15CST04-2). J.M.V.L. acknowledges funding support from NWO START-UP from the Netherlands. S.T. acknowledges funding by the Computational Sciences for Energy Research (CSER) tenure track program of Shell and NWO (Project No. 15CST04-2) as well as the NWO START-UP from the Netherlands.

# Supporting Information

## The Role of Solvents in the Formation of Methylammonium Lead Triiodide Perovskite

Junke Jiang,[1,2] José Manuel Vicent-Luna,[1,2] and Shuxia Tao,*[,1,2]

[1]Materials Simulation and Modelling, Department of Applied Physics, Eindhoven University of Technology, 5600MB Eindhoven, The Netherlands

[2]Center for Computational Energy Research, Department of Applied Physics, Eindhoven University of Technology, Eindhoven 5600 MB, The Netherlands

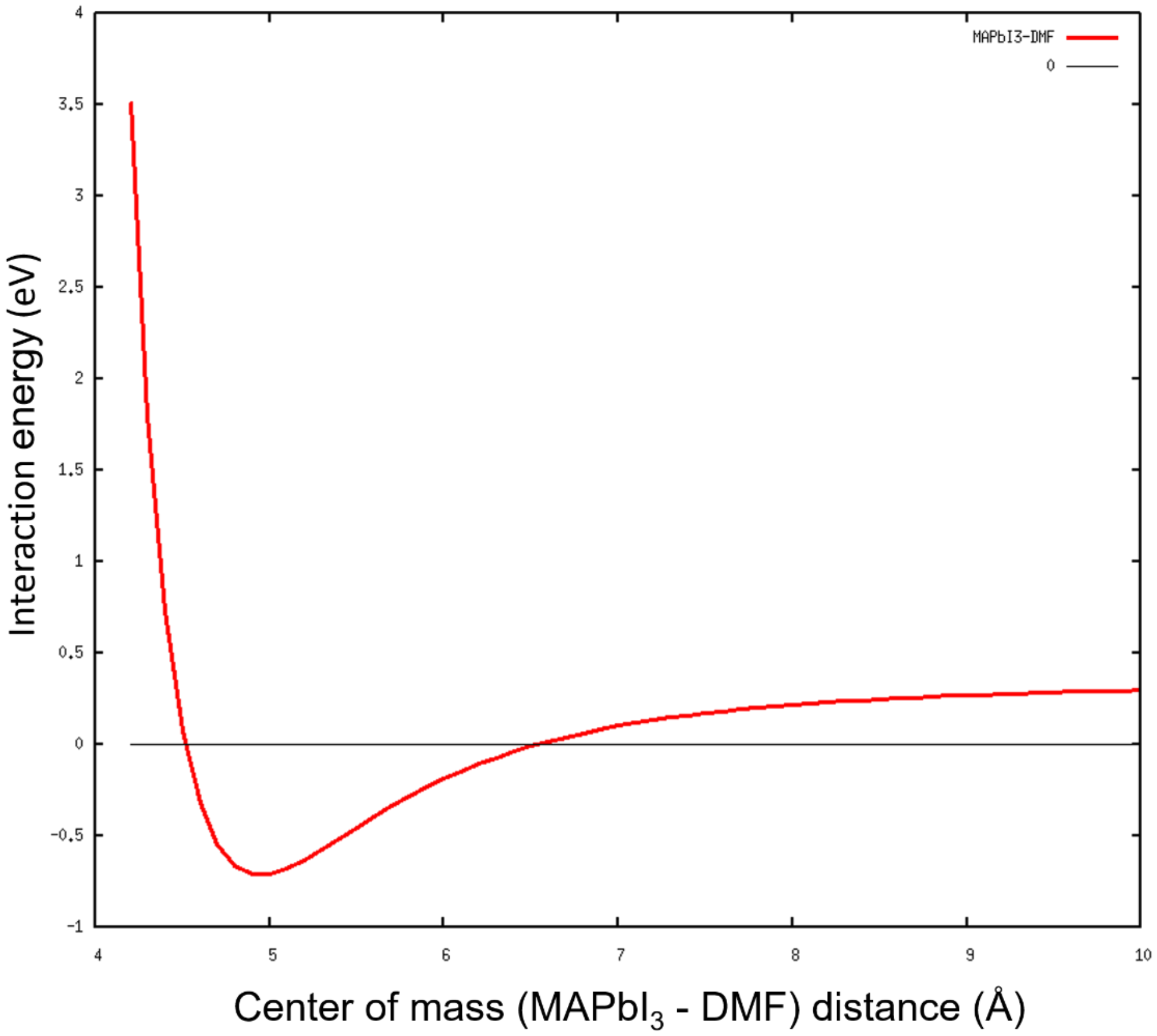

Figure S1. Interaction energies of $MAPbI_3$-DMF iodoplumbate complexes as a function of the distance of their center of masses.

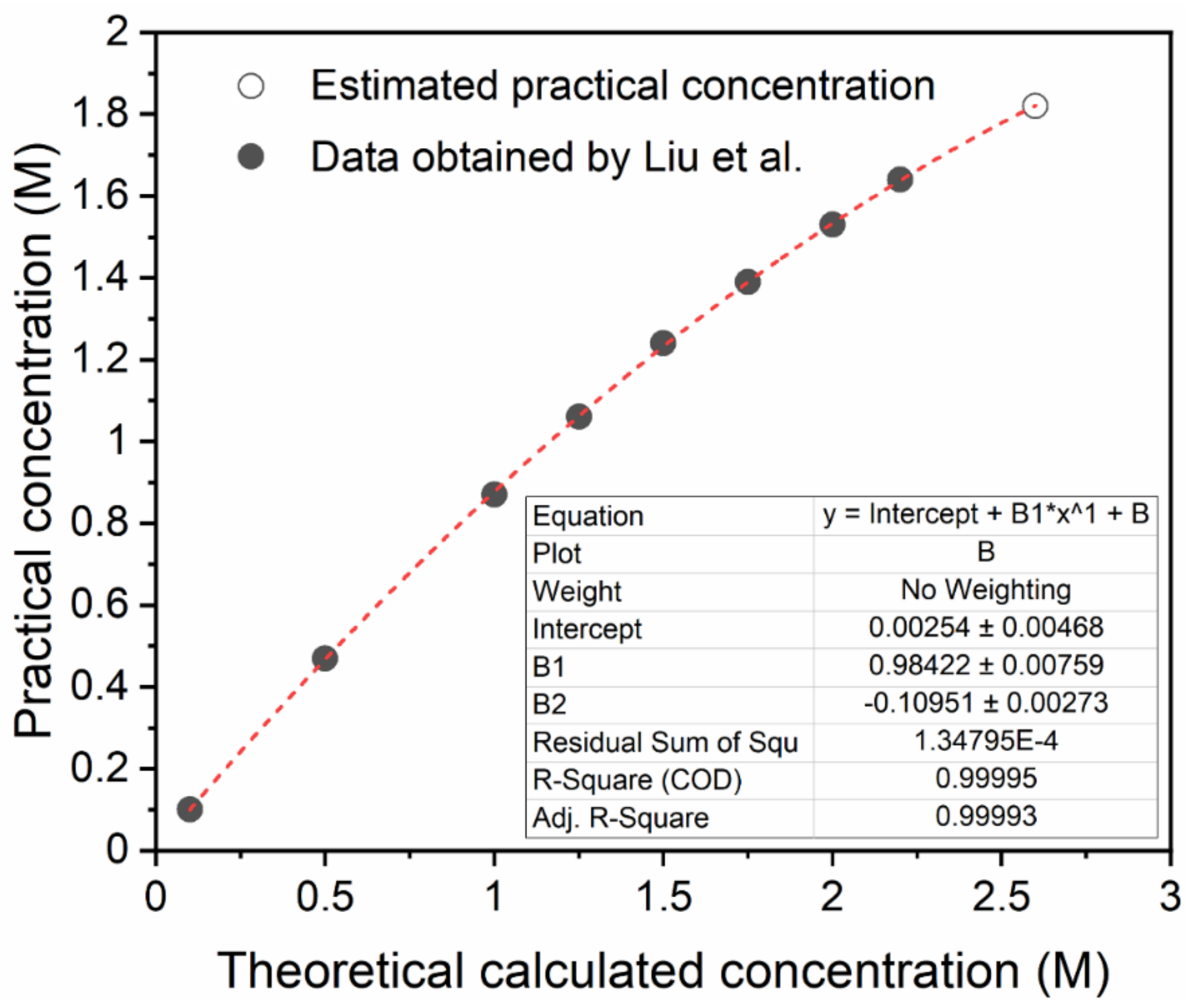

Figure S2. Correlation between theoretically calculated concentration and practical concentration. The solid dots are obtained from the results measured by Liu. et al.[1] The hollowed dot is the theoretical concentration using in this study for DMF solvent. The practical concentration is estimated as 1.82 M by fitting with the results from Ref. [1].

Table S1. Donor number and boiling point of DMSO and DMF. Values are obtained from Ref 2.

| | DMSO | DMF |
|---|---|---|
| | 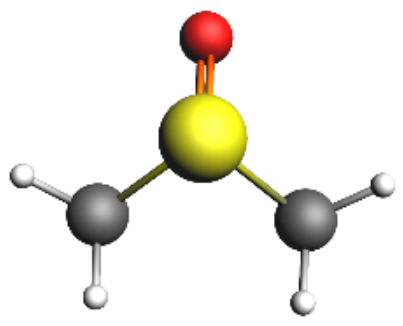 | 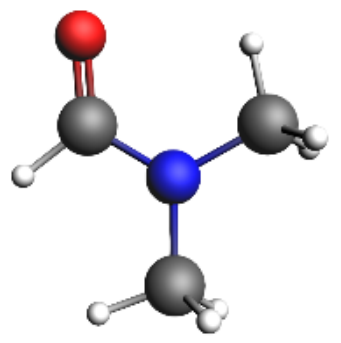 |
| Donor number | 29.8 | 26.6 |
| Boiling point (°C) | 189 | 153 |

Table S2. The step-by-step formation enthalpy of different of DMF.

| Configurations / Reactions | Formation enthalpy (eV) | | | | |
|---|---|---|---|---|---|
| | 1-1 | 2-2 | 1-2 | 2-1 | average |
| $[PbI_2L_6]$ to $[PbI_3L_6]^{1-}$ | -0.30 | -0.20 | -0.37 | -0.13 | -0.25 |
| $[PbI_3L_6]^{1-}$ to $[PbI_4L_6]^{2-}$ | -0.53 | -0.91 | -0.70 | -0.74 | -0.72 |
| $[PbI_4L_6]^{2-}$ to $[PbI_5L_6]^{3-}$ | -0.18 | -0.23 | 0.02 | -0.44 | -0.21 |
| $[PbI_5L_6]^{3-}$ to $[PbI_6L_6]^{4-}$ | -0.29 | -0.24 | -0.04 | -0.50 | -0.27 |

Table S3. The formation enthalpy of polymeric iodoplumbates of DMF.

| Configurations / Reactions | Formation enthalpy (eV) | | | | |
|---|---|---|---|---|---|
| | 1-1 | 2-2 | 1-2 | 2-1 | average |
| $[PbI_2L_4]$ to $[PbI_2L_4]_2$ | -0.33 | -0.17 | -0.24 | -0.08 | -0.20 |
| $[PbI_3L_3]^{1-}$ to $([PbI_3L_3]^{1-})_2$ | -0.03 | 0.15 | -0.21 | -0.03 | -0.03 |
| $[PbI_4L_2]^{2-}$ to $([PbI_4L_2]^{2-})_2$ | -0.46 | -0.51 | -0.30 | -0.35 | -0.41 |

Table S4. The enthalpy of two considered configurations for DMF.

| Configurations / Iodoplumbates | Enthalpy (eV) | | |
|---|---|---|---|
| | 1 | 2 | average |
| $[PbI_2L_6]$ | -397.27 | -397.20 | -397.24 |
| $[PbI_3L_6]^{1-}$ | -436.93 | -436.75 | -436.84 |
| $[PbI_4L_6]^{2-}$ | -476.81 | -477.01 | -476.91 |
| $[PbI_5L_6]^{3-}$ | -516.34 | -516.59 | -516.47 |
| $[PbI_6L_6]^{4-}$ | -555.98 | -556.18 | -556.08 |

Table S5. The step-by-step formation enthalpy of different of DMSO.

| Configurations / Reactions | Formation enthalpy (eV) | | | | |
|---|---|---|---|---|---|
| | 1-1 | 2-2 | 1-2 | 2-1 | average |
| $[PbI_2L_6]$ to $[PbI_3L_6]^{1-}$ | -0.36 | -0.46 | -0.51 | -0.31 | -0.41 |
| $[PbI_3L_6]^{1-}$ to $[PbI_4L_6]^{2-}$ | -0.46 | -0.49 | -0.51 | -0.44 | -0.48 |
| $[PbI_4L_6]^{2-}$ to $[PbI_5L_6]^{3-}$ | -0.10 | -0.05 | -0.12 | -0.03 | -0.07 |
| $[PbI_5L_6]^{3-}$ to $[PbI_6L_6]^{4-}$ | -0.71 | -0.31 | -0.78 | -0.24 | -0.51 |

Table S6. The formation enthalpy of polymeric iodoplumbates of DMSO.

| Configurations / Reactions | Formation enthalpy (eV) | | | | |
|---|---|---|---|---|---|
| | 1-1 | 2-2 | 1-2 | 2-1 | average |
| $[PbI_2L_4]$ to $[PbI_2L_4]_2$ | -0.10 | -0.04 | -0.07 | -0.01 | -0.05 |
| $[PbI_3L_3]^{1-}$ to $([PbI_3L_3]^{1-})_2$ | 0.50 | 0.48 | 0.30 | 0.28 | 0.39 |
| $[PbI_4L_2]^{2-}$ to $([PbI_4L_2]^{2-})_2$ | -0.26 | -0.15 | -0.40 | -0.29 | -0.28 |

Table S7. The enthalpy of two considered configurations for DMSO.

| Configurations / Iodoplumbates | Enthalpy (eV) | | |
|---|---|---|---|
| | 1 | 2 | average |
| $[PbI_2L_6]$ | -292.98 | -292.83 | -292.91 |
| $[PbI_3L_6]^{1-}$ | -332.73 | -332.68 | -332.70 |
| $[PbI_4L_6]^{2-}$ | -372.57 | -372.56 | -372.57 |
| $[PbI_5L_6]^{3-}$ | -412.06 | -411.99 | -412.03 |
| $[PbI_6L_6]^{4-}$ | -452.16 | -451.69 | -451.92 |

Table S8. The step-by-step formation enthalpy of different of DMF-DMSO binary.

| Configurations / Reactions | Formation enthalpy (eV) | | | | |
|---|---|---|---|---|---|
| | 1-1 | 2-2 | 1-2 | 2-1 | average |
| $[PbI_2L_6]$ to $[PbI_3L_6]^{1-}$ | -0.60 | -0.47 | -0.52 | -0.55 | -0.53 |
| $[PbI_3L_6]^{1-}$ to $[PbI_4L_6]^{2-}$ | -0.15 | -0.24 | -0.20 | -0.19 | -0.19 |
| $[PbI_4L_6]^{2-}$ to $[PbI_5L_6]^{3-}$ | -0.69 | -0.25 | -0.64 | -0.29 | -0.47 |
| $[PbI_5L_6]^{3-}$ to $[PbI_6L_6]^{4-}$ | -0.21 | -0.67 | -0.60 | -0.28 | -0.44 |

Table S9. The formation enthalpy of polymeric iodoplumbates of DMF-DMSO binary.

| Configurations / Reactions | Formation enthalpy (eV) | | | | |
|---|---|---|---|---|---|
| | 1-1 | 2-2 | 1-2 | 2-1 | average |
| $[PbI_2L_6]$ to $[PbI_2L_6]_2$ | -0.23 | -0.12 | -0.22 | -0.11 | -0.17 |
| $[PbI_3L_3]^{1-}$ to $([PbI_3L_3]^{1-})_2$ | -0.27 | -0.09 | -0.12 | 0.06 | -0.10 |
| $[PbI_4L_2]^{2-}$ to $([PbI_4L_2]^{2-})_2$ | -0.25 | -0.21 | -0.41 | -0.36 | -0.36 |

Table S10. The enthalpy of two considered configurations for DMF-DMSO binary.

| Configurations / Iodoplumbates | Enthalpy (eV) | | |
|---|---|---|---|
| | 1 | 2 | average |
| $[PbI_2L_6]$ | -379.76 | -379.84 | -379.80 |
| $[PbI_3L_6]^{1-}$ | -419.71 | -419.66 | -419.68 |
| $[PbI_4L_6]^{2-}$ | -459.20 | -459.25 | -459.23 |
| $[PbI_5L_6]^{3-}$ | -499.24 | -498.85 | -499.04 |
| $[PbI_6L_6]^{4-}$ | -538.80 | -538.87 | -538.83 |

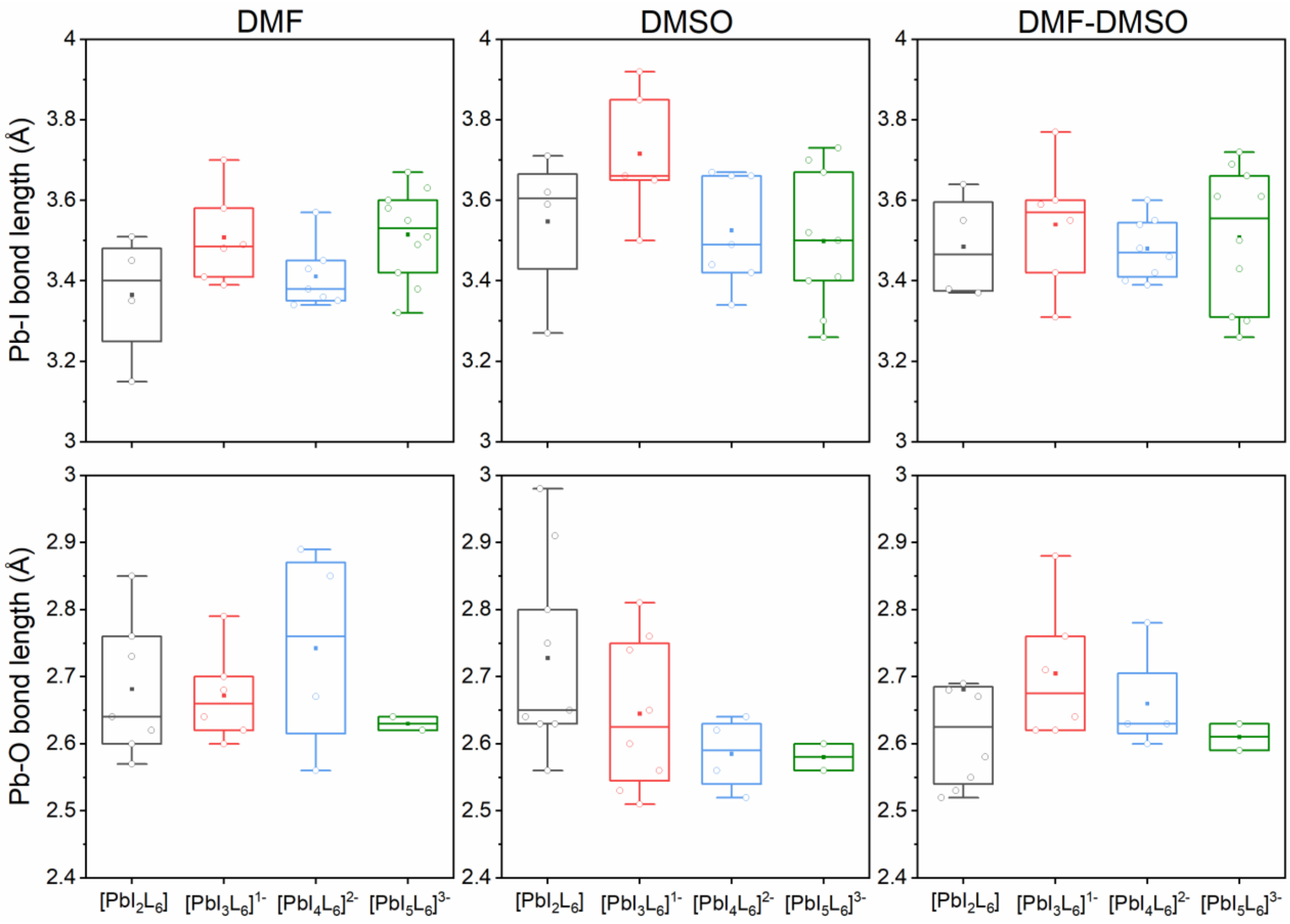

Figure S3. Pb-I and (b) Pb-O bond lengths of $[PbI_2L_6]$, $[PbI_3L_6]^{1-}$, $[PbI_4L_6]^{2-}$, and $[PbI_5L_6]^{3-}$ of DMF, DMSO, and DMF-DMSO binary, respectively.

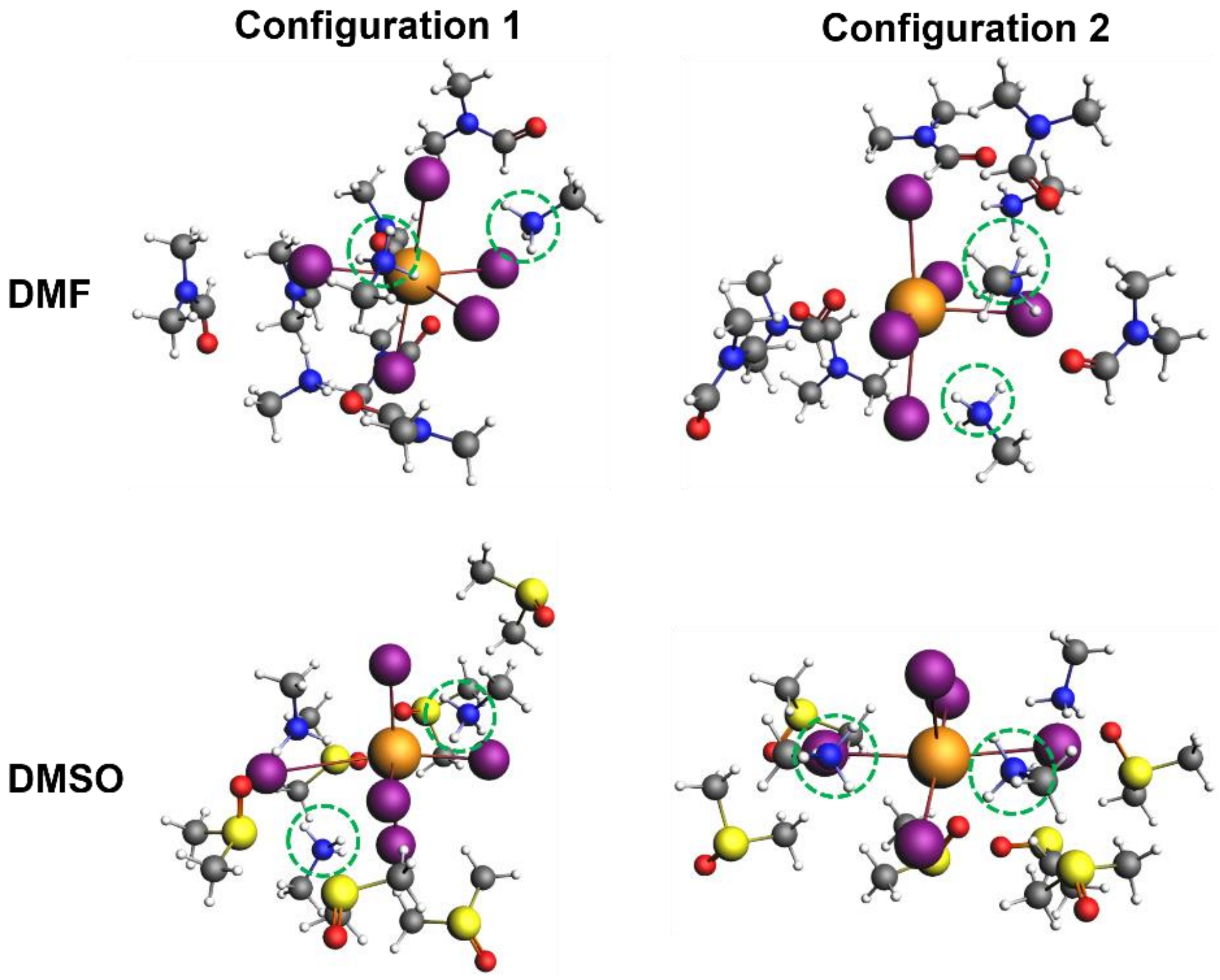

Figure S4. The optimized iodoplumbates configurations $[PbI_5L_6]^{3-}$ of DMF and DMSO, respectively. There are more NH···I hydrogen bond formed (indicated by dashed circles).

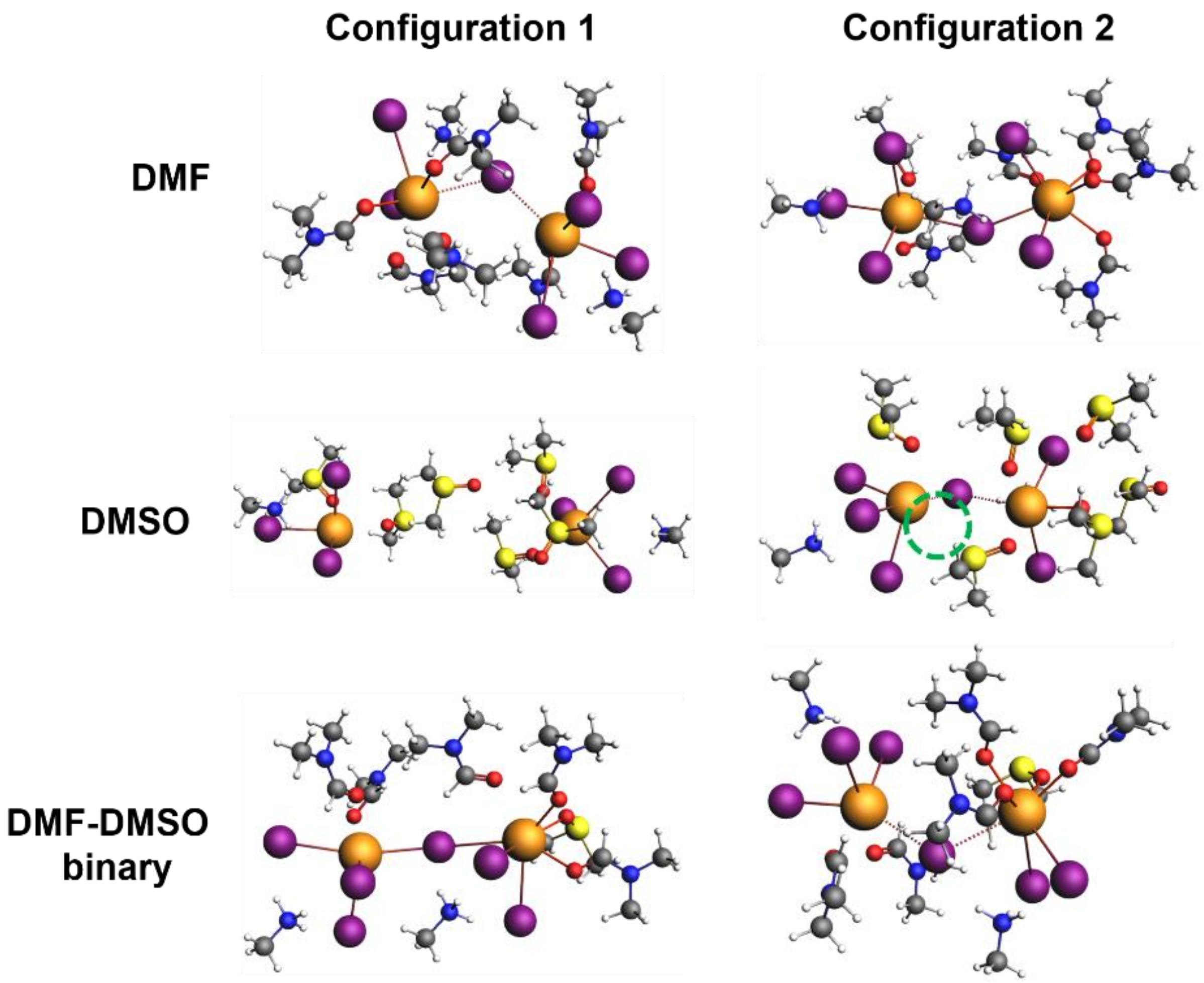

Figure S5. The optimized polymeric iodoplumbates (dimer) configurations of $[PbI_2L_6]$, $[PbI_3L_6]^{1-}$, $[PbI_4L_6]^{2-}$, and $[PbI_5L_6]^{3-}$ of DMF, DMSO, and DMF-DMSO binary, respectively. The $([PbI_3L_3]^{1-})_2$ are not formed (left), or not fully bonded with solvents (indicated by a dashed circle).

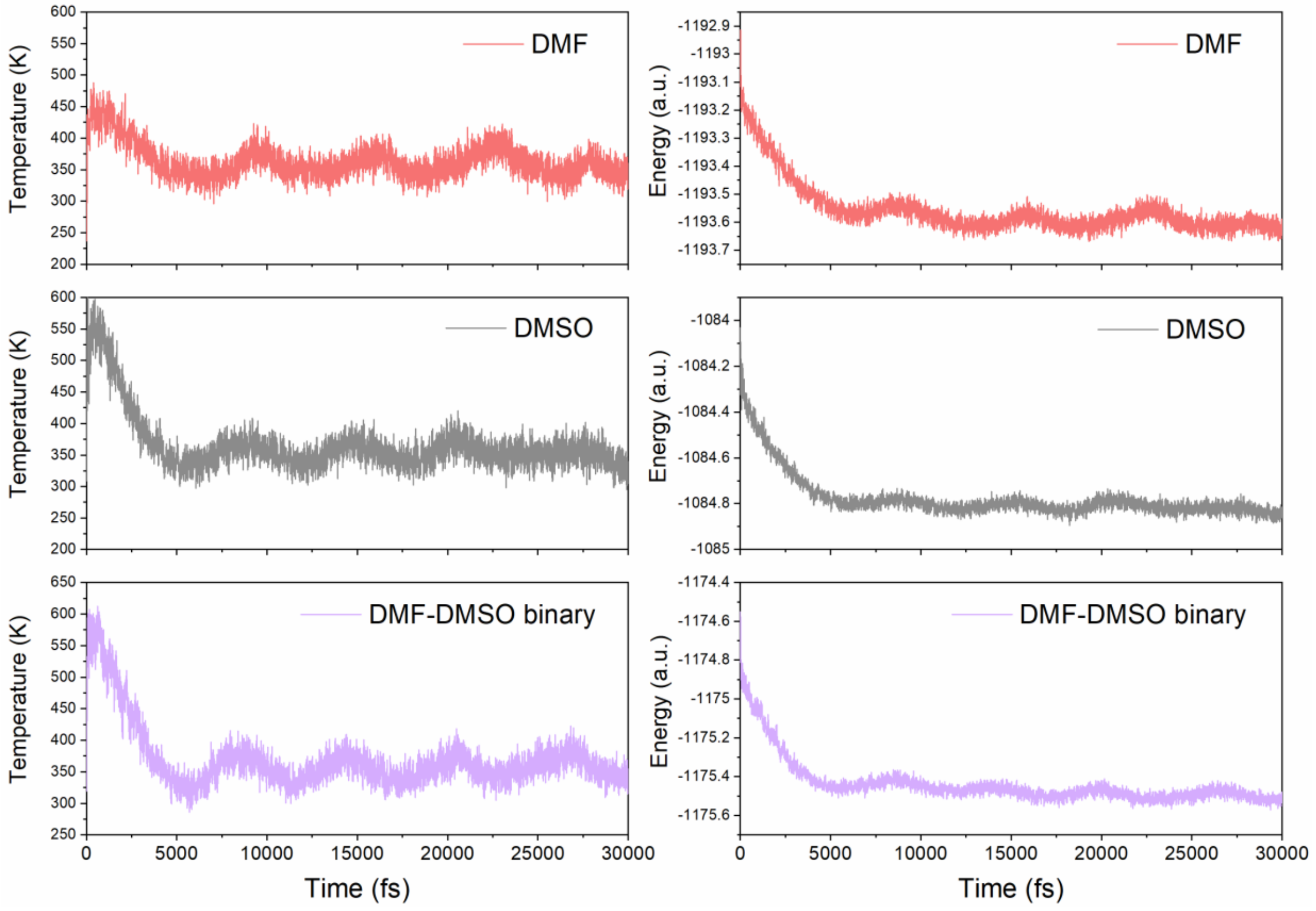

Figure S6. These temperature and energy oscillations show that the simulations reach an equilibrium after about 5000 fs. Therefore, the trajectory of the first 5000 fs in each simulation is depleted for the data analysis.

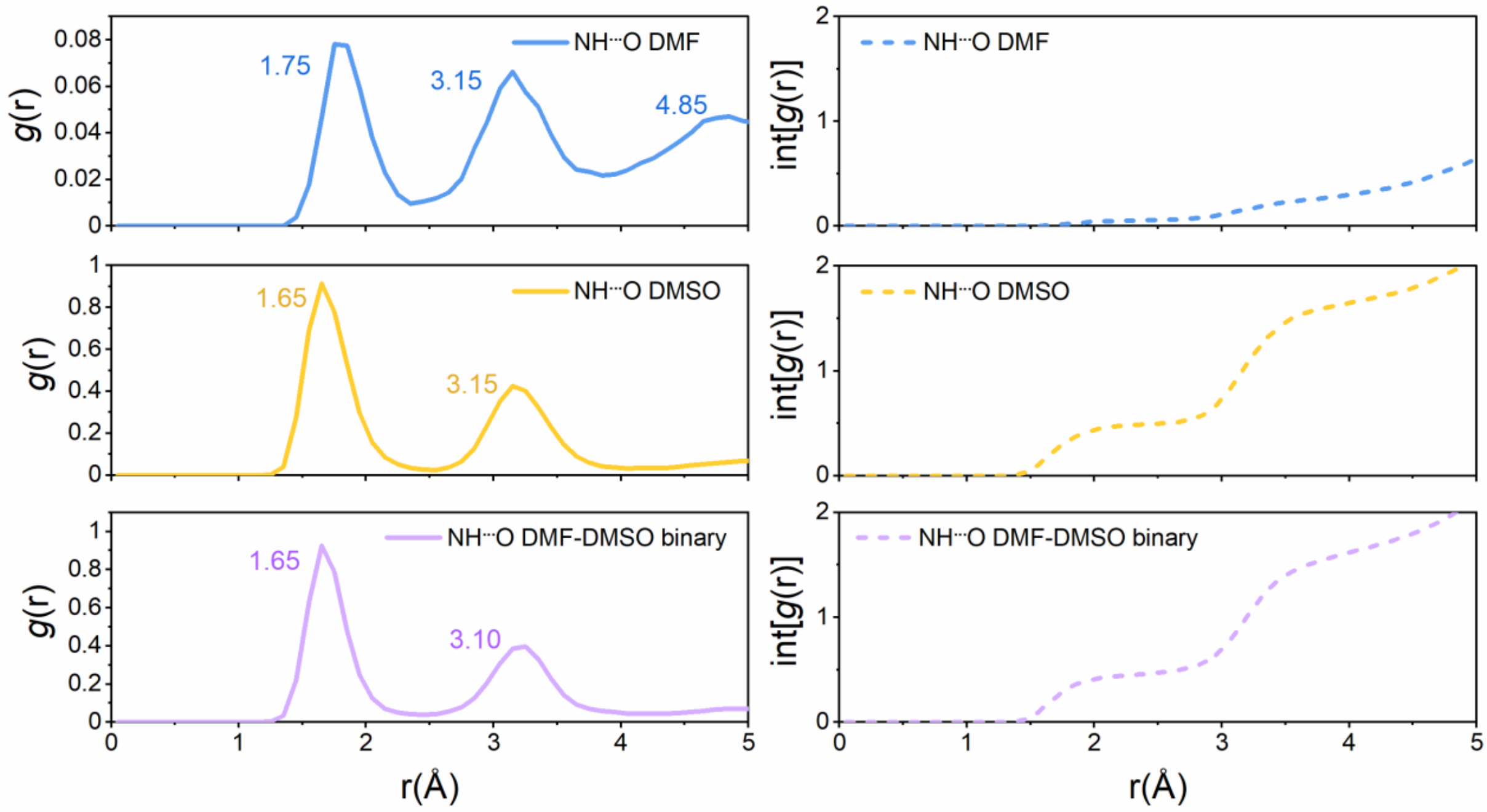

Figure S7. Short-range radial distribution function *g*(r) (left) and cumulative radial distribution functions int[*g*(r)] (right) of NH···O (hydrogen bond between $MA^+$ and solvents) for DMF, DMSO, and DMF-DMSO binary solvents, respectively.